\documentclass[twocolumn,showpacs,prl,aps]{revtex4-1}
\usepackage{bm,color,amsmath,amssymb,mathrsfs,latexsym,graphicx,psfrag}

\newcommand{\bs}[1]{\boldsymbol{#1}}
\newcommand{\be}{\begin{equation}}
\newcommand{\ee}{\end{equation}}
\newcommand{\bea}{\begin{eqnarray}}
\newcommand{\eea}{\end{eqnarray}}

\renewcommand{\phi}{\varphi}
\renewcommand{\epsilon}{\varepsilon}

\begin{document}

\title{Competing many-body instabilities and unconventional superconductivity in graphene}

\author{Maximilian Kiesel${}^1$} 
\author{Christian Platt${}^1$} 
\author{Werner Hanke${}^1$} 
\author{Dmitry A. Abanin${}^{2}$}
\author{Ronny Thomale${}^{1,3}$}
\affiliation{${}^1$Institute for Theoretical
  Physics, University of W\"urzburg, Am Hubland, D
  97074 W\"urzburg}
\affiliation{${}^2$Department of Physics, Harvard University, Cambridge, MA 02138, USA} 
\affiliation{${}^3$Department of Physics, Stanford University, Stanford, California 94305, USA}

\date{\today}

\begin{abstract}
The band structure of graphene exhibits van Hove singularities (VHS) at doping $x=\pm 1/8$ away from the Dirac point. Near the VHS, interactions effects, enhanced due to the large density of states, can give rise to various many-body phases at experimentally accessible temperatures. We study the competition between different many-body instabilities in graphene using functional renormalization group (FRG). We predict a rich phase diagram, which, depending on long range hopping as well as screening strength and absolute scale of the Coulomb interaction, contains a $d+id$-wave superconducting (SC) phase, or a spin density wave phase at the VHS. The $d+id$ state is expected to exhibit quantized charge and spin Hall response, as well as Majorana modes bound to vortices. In the vicinity of the VHS, we find singlet $d+id$-wave as well as triplet $f$-wave SC phases.

\date{\today}

\pacs{73.22.Pr,74.70.Wz,74.20.Mn}

\end{abstract}
\maketitle

{\it Introduction.} Graphene, a monolayer of carbon, hosts a two-dimensional electronic system (2DES) with unique properties~\cite{castroneto09rmp}.  In particular, the Coulomb interaction plays an important role in graphene~\cite{kotov10arxiv}, giving rise to interesting many-body phenomena, including marginal Fermi-liquid behavior~\cite{gonzalez99prb59}, energy-dependent renormalization of the Fermi velocity~\cite{gonzalez94nuclphys424}, as well many-body states in the quantum Hall effect regime~\cite{castroneto09rmp}.

Experimentally, graphene offers a high degree of tunability. Most importantly, carrier density can be controlled in a broad range. Near the Dirac point (doping level of electrons $x=1/2$), such control is achieved by backgates and local top gates~\cite{castroneto09rmp}. Recently, it was demonstrated that chemical doping~\cite{mcchesney10prl104} and electrolytic gating~\cite{efetov10prl105} enable doping graphene far away from the Dirac point, where the band structure is no longer Dirac-like. In particular, the density can be tuned to the vicinity of the van Hove singularities (VHS) in the band structure, which occur at doping values $x=3/8,5/8$. In the case of chemical doping, the dopants form a superlattice on top of the graphene sheet. This strongly reduces the amount of disorder induced by doping. Furthermore, the spacing of the superlattice is so large that hybridization in the dopant layer can be neglected, and hence transport measurements of the graphene sample remain unaffected.

Before the strong doping of graphene has been recently accomplished experimentally, superconductivity had been predominantly studied around the Dirac point, including $p+ip$-wave from electron-phonon or plasmonic~\cite{uchoa07prl98} as well as $f$ or $d+id$-wave from electron-electron~\cite{honerkamp08prl100} interaction effects. Mean-field treatments~\cite{baska02,black07prb75} were found to be unreliable as they arrive at unrealistic $T_c>1000\, {\rm K}$, with only slightly better results for variational approaches~\cite{pathak08arxiv}. SC has not been observed experimentally in this regime, due to small electronic density of states as well as weak phonon effects~\cite{calandra05prl95}.


Near a VHS, opposed to the Dirac point regime, electronic interaction effects are expected to be strongly enhanced due to the logarithmically diverging density of states and near-nested Fermi-surface~\cite{mcchesney10prl104}. In this regime, many-body states with appreciable critical temperatures may arise. Possible candidate states include charge-density wave (CDW), a spin-density wave (SDW), or a SC state. Generally, a subtle interplay of kinetic and interaction parameters is expected to decide which many body instability is preferred at the VHS. For graphene, the additional complication arises that as the band width ($\sim 17$eV) is of the order of the interaction scale ($\sim 10$eV), graphene cannot be suitably described from the viewpoint of strict weak coupling approaches, and adopting a picture of intermediate coupling is necessary. Rephrased in terms of diagrammatic expansions starting from the non-interacting problem, this amounts to investigating the importance of leading {\it and} subleading divergent classes of diagrams. In particular, this is relevant for the competition between magnetic and SC phases in this kind of systems, one recent example of which have been the iron pnictides~\cite{asvp,maiti}.

\begin{figure}[t]
  \begin{minipage}[l]{0.99\linewidth}
    \includegraphics[width=\linewidth]{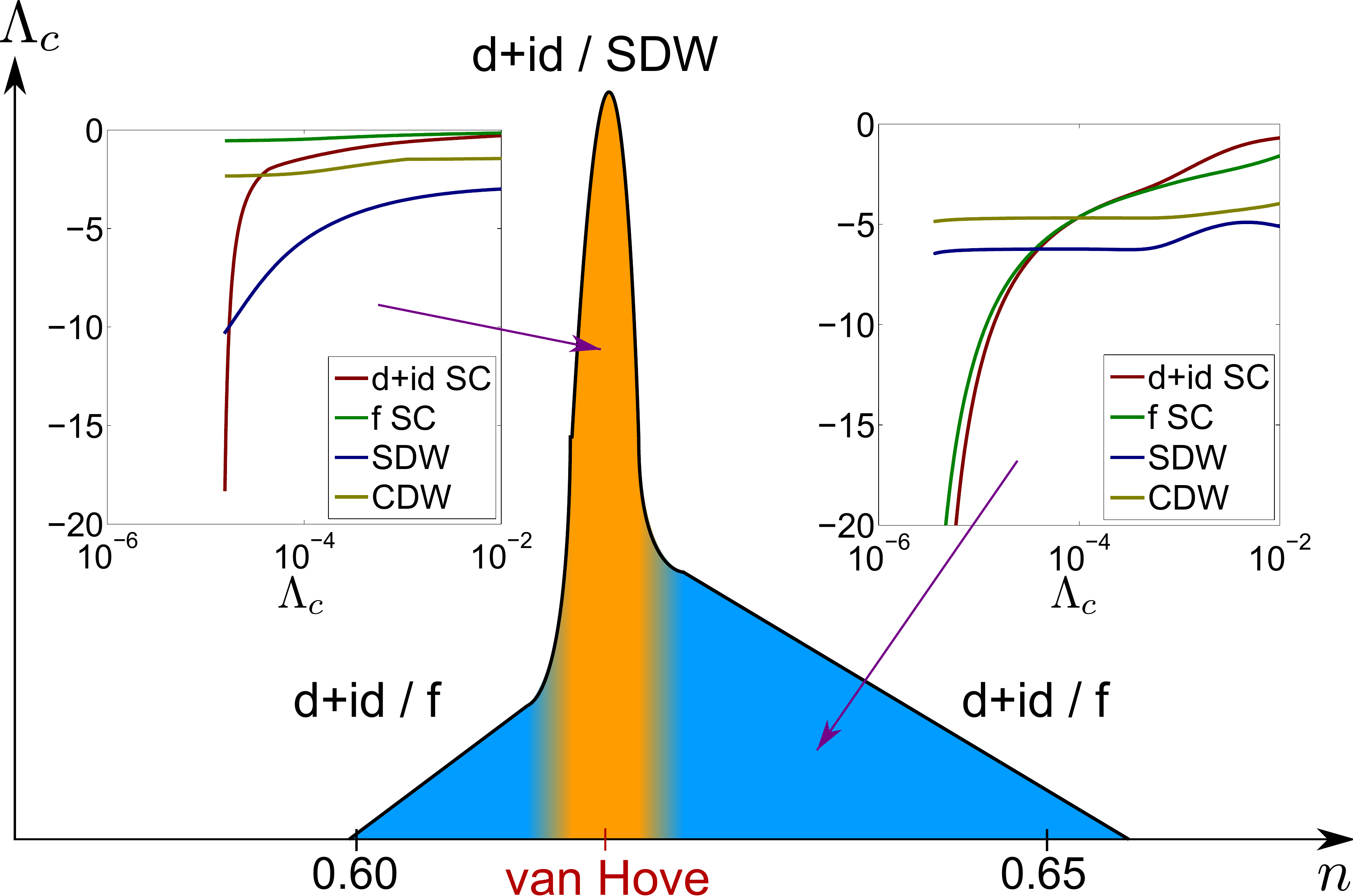}
  \end{minipage}
  \caption{(Color online). Schematic phase diagram displaying the critical instability scale $\Lambda_c \sim T_c$ as a function of doping. At the van Hove singularity (VHS, light shaded (orange) area), $d+id$ competes with spin density wave (SDW) (left flow picture: dominant $d+id$ instability for $U_0=10$eV and the band structure in~\cite{mcchesney10prl104}). Away from the VHS (dark shaded (blue) area), $\Lambda_c$ drops and whether the $d+id$ or $f$-wave SC instability is preferred depends on the long-rangedness of interaction (right flow picture: $U_1/U_0=0.45$ and $U_2/U_0=0.15$). } 
\label{pic2}
\vspace{-0pt}
\end{figure}

\begin{figure}[t]
  \begin{minipage}[l]{0.99\linewidth}
    \includegraphics[width=\linewidth]{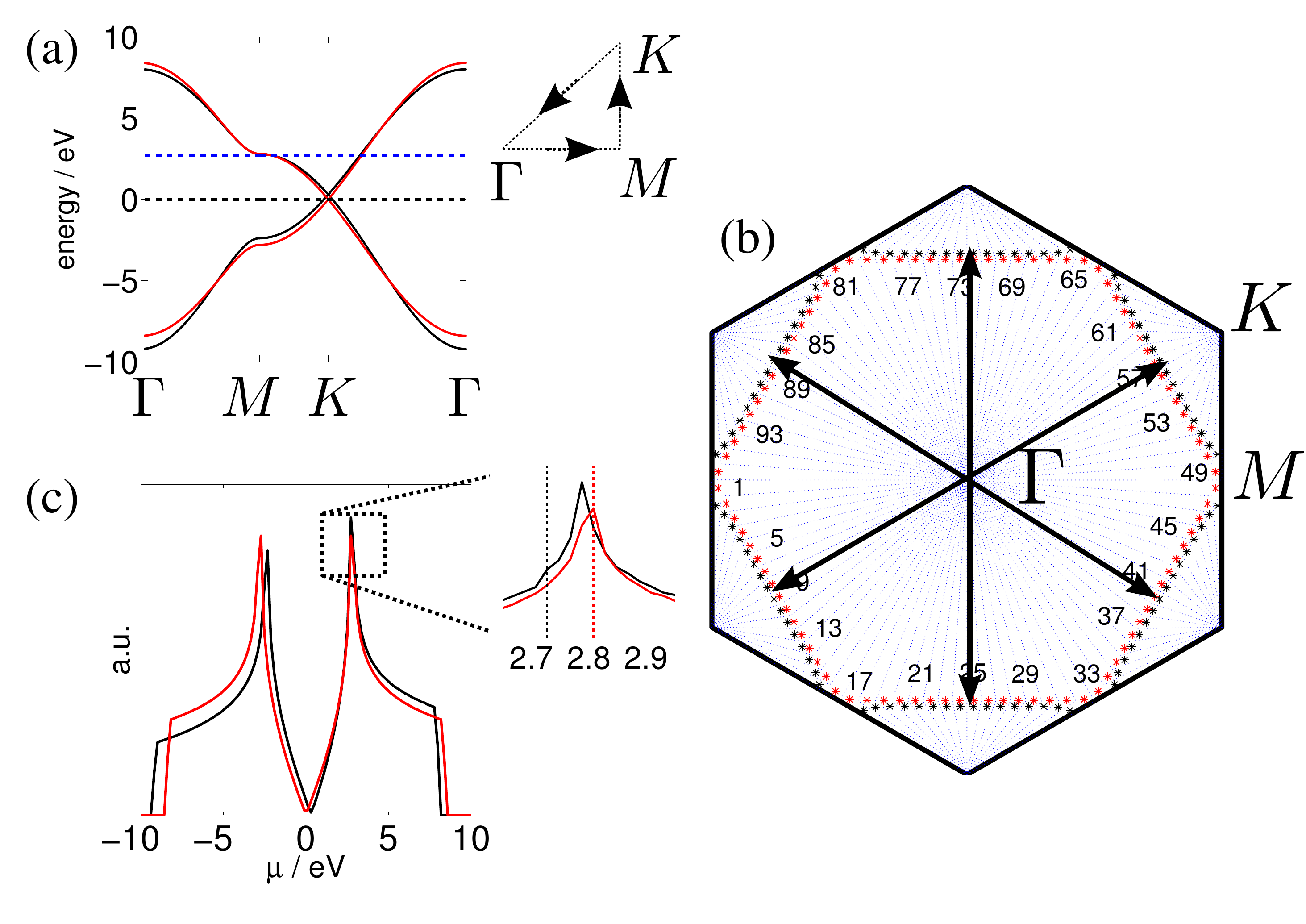}
  \end{minipage}
  \caption{(Color online). (a) Band structure of graphene once for $t_1=2.8$eV (red) and $t_1=2.8, t_2=0.7, t_3=0.02$eV (black). (b) Brillouin zone displaying the Fermi surface near the van Hove point (dashed blue level in (a), 96 patches used in the FRG and the nesting vector, and the partial nesting vectors. (c) Density of states for both band structures in (a). The inset show the position shift of Fermi surface nesting (dashed vertical lines) versus the VHS peak.}
\label{pic1}
\vspace{-0pt}
\end{figure}

{\it Main results.} In this Letter, we use the functional renormalization group (FRG) method~\cite{zanchi-00prb13609,halboth-00prb7364,honerkamp-01prb035109,honer} to study the competition between many-body states in graphene doped to the vicinity of the VHS, and attempt to analyze this problem at a level which provides a detailed connection to the experimental setup. From our analysis we obtain a rich phase diagram which, depending on the range of chosen kinetic and interaction parameters, contains magnetic and different SC phases, summarized in Fig.~\ref{pic2}.
For a certain range of parameters, we find a $d+id$ SC phase which has been previously studied by RPA~\cite{gonzalez08prb78,mcchesney10prl104}, and, very recently, perturbative three-patch renormalization group (3RG)~\cite{chubukov11arxiv}. 
To analyze all possible many-body phases and their dependence on the system parameters, FRG provides a systematic unbiased summation of diagrams in both particle-particle channels {\it and} particle-hole channels as well as vertex corrections,
 and keeps track of the whole Fermi surface [Fig.~\ref{pic1}a]. We investigate in detail how different band structure parameters affect the phase diagram. We find that rather small variations of the longer range hopping parameters such as next nearest ($t_2$) and next-next-nearest ($t_3$) hopping can shift the position of perfect Fermi surface nesting against the VHS [Fig.~\ref{pic1}], which significantly influences the competition between magnetism and SC.  
Moreover, in particular away from the exact VHS, the reduced screening of the Coulomb interaction does not justify the assumption of a local Hubbard model description. For this case, we find that only a small fraction of longer-ranged Hubbard interaction~\cite{wehling11prl106} can significantly change the phase diagram, as CDW fluctuations become more competitive to SDW fluctuations, and a triplet SC phase can appear. In particular, we study how the Cooper pairing in the different SC phases responds to differently long-ranged Hubbard interactions. Our results suggest that in experiment, modifications of the band structure as well as changing the dielectric environment of the graphene sample would enable the realization of different many-body states and possible phase transitions between them.

%

{\it Model.} We consider the $\pi$ band structure of graphene approximated by a tight binding model including up to 3rd nearest neighbors on the hexagonal lattice:
\begin{eqnarray}
H_0 &=& \Big[ t_1\sum_{\langle i,j\rangle}\sum_{\sigma} c^{\dagger}_{i,\sigma}c^{\phantom{\dagger}}_{j,\sigma} + 
      t_2\sum_{\langle\langle i,j\rangle\rangle}\sum_{\sigma} c^{\dagger}_{i,\sigma}c^{\phantom{\dagger}}_{j,\sigma}  \nonumber \\ &&
+  t_3\sum_{\langle\langle\langle i,j\rangle\rangle\rangle}\sum_{\sigma} c^{\dagger}_{i,\sigma}c^{\phantom{\dagger}}_{j,\sigma} + {\text h.c.} \Big] - \mu n,
\end{eqnarray}
where $n=\sum_{i,\sigma} n_{i,\sigma}=\sum_{i,\sigma} c_{i,\sigma}^\dagger c_{i,\sigma}^{\phantom{\dagger}}$, and $c^\dagger_{i,\sigma}$ denotes the electron annihilation operator of spin $\sigma=\uparrow, \downarrow$ at site~$i$. The resulting band structure is a two band model due to two atoms per unit cell [Fig.~\ref{pic1}]. There are certain uncertainties about the most appropriate tight binding fit for graphene, in particular as it concerns the longer range hybridization integrals~\cite{castroneto09rmp,reich}. For dominant $t_1$, the band structure features a van Hove singularity (VHS) at $x=3/8, 5/8$. Constraining ourselves to the electron-doped case, the $x=5/8$ electron-like Fermi surface is shown in Fig.~\ref{pic1}b. As depicted, this is the regime of largely enhanced density of states which we investigate in the following.  For $t_2=t_3=0$ [red curve in Fig.~\ref{pic1}], the VHS coincides with the partial nesting of different sections of the Fermi surface for $Q=(0,2\pi/\sqrt{3}), (\pi, \pi/\sqrt{3})$, and $(\pi, -\pi/\sqrt{3})$. For a realistic band structure estimate with finite $t_2$ and $t_3$~\cite{mcchesney10prl104} [black curve in Fig.~\ref{pic1}], this gives a relevant shift of the perfect nesting position versus the VHS as well as density of states at the VHS, and affects the many-body phase found there.

We assume Coulomb interactions represented by a long range Hubbard Hamiltonian~\cite{wehling11prl106}
\begin{eqnarray}
H_{\text{int}}&=& U_{0} \sum_i n_{i, \uparrow} n_{i, \downarrow} + \frac{1}{2} U_{1} \sum_{\langle i,j\rangle, \sigma, \sigma'} n_{i, \sigma} n_{j, \sigma'}\nonumber \\
&&+ \frac{1}{2} U_{2} \sum_{\langle \langle i,j\rangle \rangle, \sigma, \sigma'} n_{i, \sigma} n_{j, \sigma'},
\end{eqnarray}
where $U_{0\dots 2}$ parametrizes the Coulomb repulsion scale from onsite to the second nearest neighbor interaction. It depends on the density of states how strongly the Coulomb interaction is screened. At the VHS, we assume perfect screening and consider $U_0$ only, while away from the VHS, we investigate the phenomenology of taking $U_1$ and $U_2$ into consideration. The typical scale of the effective $U_0$ has been found to be $\sim 10 \text{eV} < W$~\cite{wehling11prl106}, where $W\sim 17 \text{eV}$ is the kinetic bandwidth.

{\it Method.} We employ the FRG and study how the renormalized interaction described by the 4-point
function (4PF) evolves under integrating high energy fermionic modes:
$V_{\Lambda}(\bs{k}_1;\bs{k}_2;\bs{k}_3;\bs{k}_4)c_{\bs{k}_4s}^{\dagger}c_{\bs{k}_3\bar{s}}^{\dagger}c_{\bs{k}_2s}^{\phantom{\dagger}}c_{\bs{k}_1\bar{s}}^{\phantom{\dagger}},$
where the flow parameter is the IR cutoff $\Lambda$ approaching the
Fermi surface, and with $\bs{k}_{1}$ to $\bs{k}_{4}$ the incoming and
outgoing momenta. Within the numerical treatment, the $\bs{k}$'s are discretized to take on the values representing the different patches of the Brillouin zone. Fig.~\ref{pic1}b shows a 96 patching scheme. We checked for selected representative scenarios that our results are converged against supercomputer simulations with 192 patch resolution.
 The starting conditions of the RG
are given by the bandwidth $W$ serving as an UV cutoff, with the bare
initial interactions for the 4PF. 
Due to the spin rotational invariance of
interactions (we neglect spin-orbit coupling in our analysis), we constrain ourselves to the $S^z=0$ subspace of
incoming momenta $\bs{k}_1, \bs{k}_2$ (and outgoing
$\bs{k}_3,\bs{k}_4$) and generate the singlet and triplet channel by
symmetrization and antisymmetrization of the 4PF
$V_{\Lambda}$~\cite{honerkamp-01prb035109}. 
The diverging channels of the 4PF
under the flow to the Fermi surface signal the nature of the
instability, and the corresponding $\Lambda_c$, as a function of some given system parameter such as doping, gives the same qualitative behavior as $T_c$. At a cutoff scale where the leading instability starts to diverge, we then decompose the different channels such as SC or SDW into different eigenmode contributions and obtain the form factors associated with the different instabilities~\cite{zhai-09prb064517,asvp}. 

{\it Phase diagram.} The phase diagram as a function of doping, obtained for realistic microscopic kinetic and interaction parameters~\cite{mcchesney10prl104,wehling11prl106} is shown in Fig.~\ref{pic2}.  
At the VHS [orange-shaded area in Fig.~\ref{pic2}], the density of states is so large that a local Hubbard description is appropriate. For $U_0 \sim 10 eV$, we find that the $d+id$ SC instability is dominant, assuming finite hopping parameters $t_2=0.7$eV and $t_3=0.02$eV~\cite{mcchesney10prl104}. (The result is rather similar for the values of~\cite{reich}.) Only at scales of $U_0>18$ eV the SDW becomes dominant for this scenario. Note, however, that only small modifications of the band structure can strongly affect the competition of SDW and SC at the VHS: when $t_2$ is reduced, the system gets more biased to the SDW, as the SDW fluctuations in the nesting channel get enhanced. For $t_1$ only [red curve in Fig.~\ref{pic1}a], the SDW already wins for $U_0> 8.5$ eV.  As we move away from the VHS [blue-shaded area in Fig.~\ref{pic2}], details of the band structure become less relevant, and we note that the critical instability scale $\Lambda_c$ drops stronger towards the Dirac point than away from it, mainly as a consequence of the reduced density of states. As SDW fluctuations are weakened, SC phases become dominant. Still assuming rather local Coulomb interactions ($U_1/U_0 < 0.40 $), we find that the system still favors the $d+id$ SC state. Allowing for more long-ranged Hubbard interactions, however, the picture changes: the CDW fluctuations are comparable to the SDW fluctuations which would bias the system towards singlet SC, and  triplet $f$-wave pairing becomes competitive.

\begin{figure}[t]
  \begin{minipage}[l]{0.99\linewidth}
    \includegraphics[width=\linewidth]{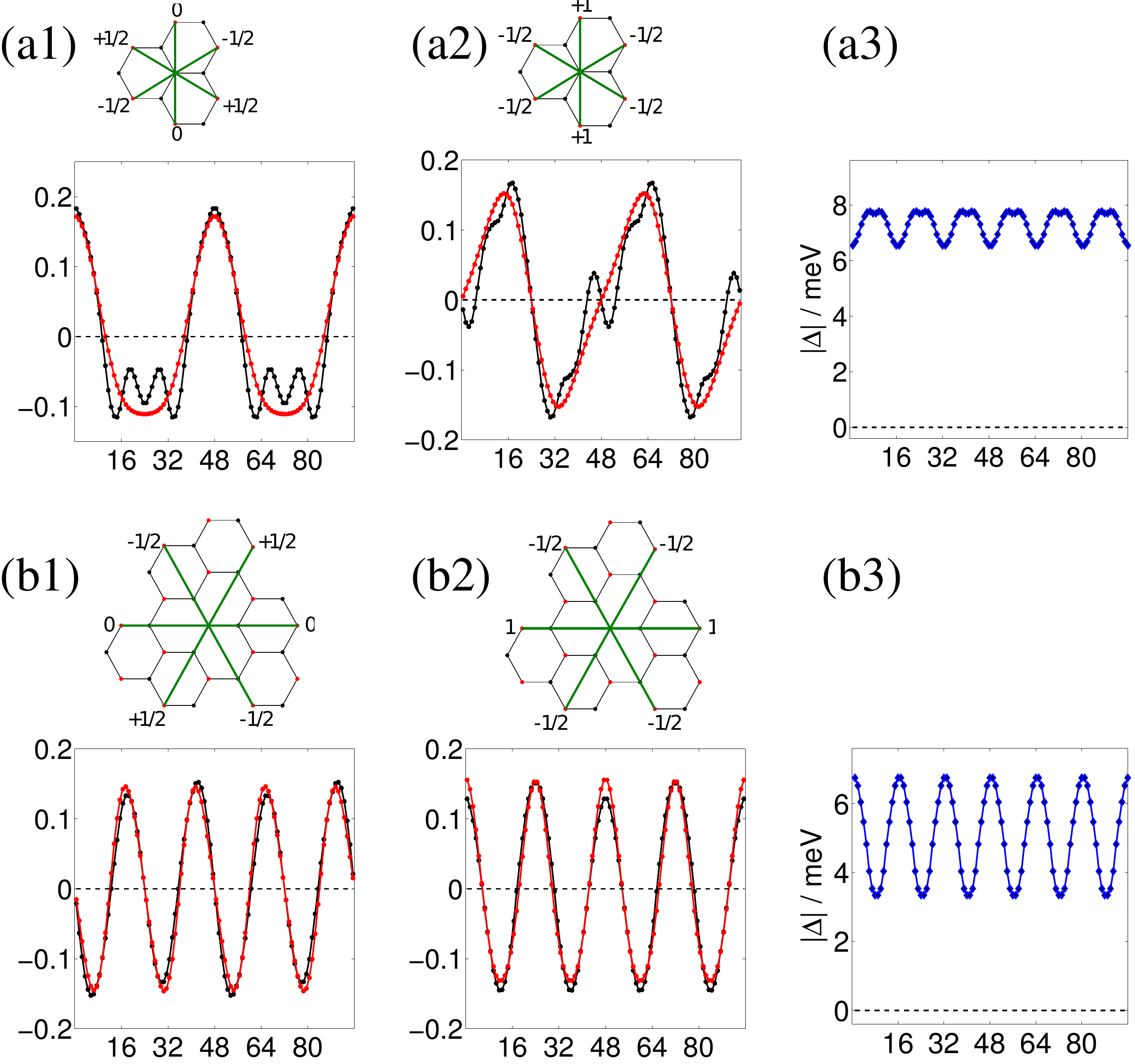}
  \end{minipage}
  \caption{(Color online). $d_{x^2-y^2}$ and $d_{xy}$-wave solutions for (a) $U_0=10$eV only (and (b) $U_1/U_0=0.4$. (a1), (a2) and (b1), (b2)  show the form factors of $d_{x^2-y^2}$ and $d_{xy}$ plotted along the Fermi surface according to patch indices defined in Fig.~\ref{pic1}b, as well as the real space pair amplitude patterns. The solutions change from (a) to (b). The analytic form factors given in the text (red) fit the numerical data (black). (a3) and (b3) show the gap profile of $d+id$ along the Fermi surface (actual connection to experimental energy scale can still vary by a global factor). The gap anisotropy increases from (a) to (b).}
\label{pic3}
\vspace{-0pt}
\end{figure}

{\it $d+id$-wave phase.} Let us analyze the $d$-wave SC phase at the VHS ($U_0$ only) in more detail. The honeycomb lattice is characterized by $C_{6v}$ symmetry about the center of hexagons, and the SC order parameter transforms as one of the irreducible representations. $d_{x^2-y^2}$ and $d_{xy}$-wave follow the two-dimensional $E_2$ representation and are hence degenerate.  The different form factors are plotted in Fig.~\ref{pic3}a. We find that the numerical solutions can be fit to $f[d_{x^2-y^2}]=2\cos(\sqrt{3}k_y) - \cos[(\sqrt{3}k_y - 3k_x)/2] -  \cos[(\sqrt{3}k_y + 3k_x)/2]$ and $f[d_{xy}]=\cos[(\sqrt{3}k_y - 3k_x)/2] -  \cos[(\sqrt{3}k_y + 3k_x)/2]$. From the Fourier transform of the momentum-resolved form factors $f(\bs{k})$ along the Fermi surface we also obtain the pairing amplitudes of the real space SC pairing function~\cite{saheb} [Fig.~\ref{pic3}]. The Cooper pairing emerges on nearest neighbors of the same hexagonal sublattice. As we move to the broader vicinity of the VHS where we assume longer range Hubbard interaction, the form factors retain the $d$-wave $E_2$ representation, while the Cooper pair wave function changes as shown in Fig.~\ref{pic3}b ($U_1/U_0=0.4$, $U_2/U_0=0.25$). There, the form factor fits change to $f[d_{x^2-y^2}]=2\cos(3k_x) - \cos[(3\sqrt{3}k_y - 3k_x)/2] -  \cos[(3\sqrt{3}k_y + 3k_x)/2]$ and $f[d_{xy}]=\cos[(3\sqrt{3}k_y - 3k_x)/2] -  \cos[(3\sqrt{3}k_y + 3k_x)/2]$, corresponding to a doubled number of nodes along the Fermi surface. From the pairing amplitudes, we likewise observe that the pairing spreads out to the second nearest neighbor of the same sublattice. This is a consequence of the long-range Hubbard interactions: the Cooper pair wave function seeks to develop more nodes to minimize Coulomb repulsion, and is able to do so by longer range Cooper pairing. This, however, still does not tell us about the gap function of the $d$-wave instability. As the degeneracy is protected by symmetry, the system could generically form any linear combination $ d_{x^2-y^2}+e^{i\theta}d_{xy}$ of both $d$-wave solutions. For this purpose, we perform a mean field decoupling in the SC pairing channel and minimize the free energy as a function of the superposition parameter. The necessary condition for such a minimum can be equivalently rephrased by satisfying the self-consistent gap equation~\cite{spid}
\begin{equation}
\Delta_{\bs{q}}=-1/N\sum_{\bs{k}}V^{\text{SC}}(\bs{k},\bs{q})\frac{\Delta_{\bs{k}}}{2 E(\bs{k})} \text{tanh} \left( \frac{E(\bs{k})}{2T}\right).
\end{equation}
The gap functions are displayed in Figs.~\ref{pic3}(a3) and~\ref{pic3}(b3). We always find $d+id$ to be the energetically preferred combination. This is rather generic in a situation of degenerate nodal SC order parameters, since such a combination allows the system to avoid nodes in the gap function. The gaps we find are hence nodeless and only slightly change their anisotropy as the pairing function varies [Fig.~\ref{pic3}a and~\ref{pic3}b].
 As graphene can be tuned rather accurately to the van Hove filling where we find the highest critical scale, it may be a reasonably accesible experimental system to study such a SC phase. The expected experimental evidence for $d+id$ would hence be a nodeless gap detectable through transport measurements and singlet pairing due to a Knight shift drop below $T_c$.  A minor caveat is given by the role of impurities which may spoil the symmetry between the two $d$-wave solutions, which could give rise to a nodal gap beyond sufficient impurity concentration~\cite{matzeflorens}. 

\begin{figure}[t]
  \begin{minipage}[l]{0.99\linewidth}
    \includegraphics[width=\linewidth]{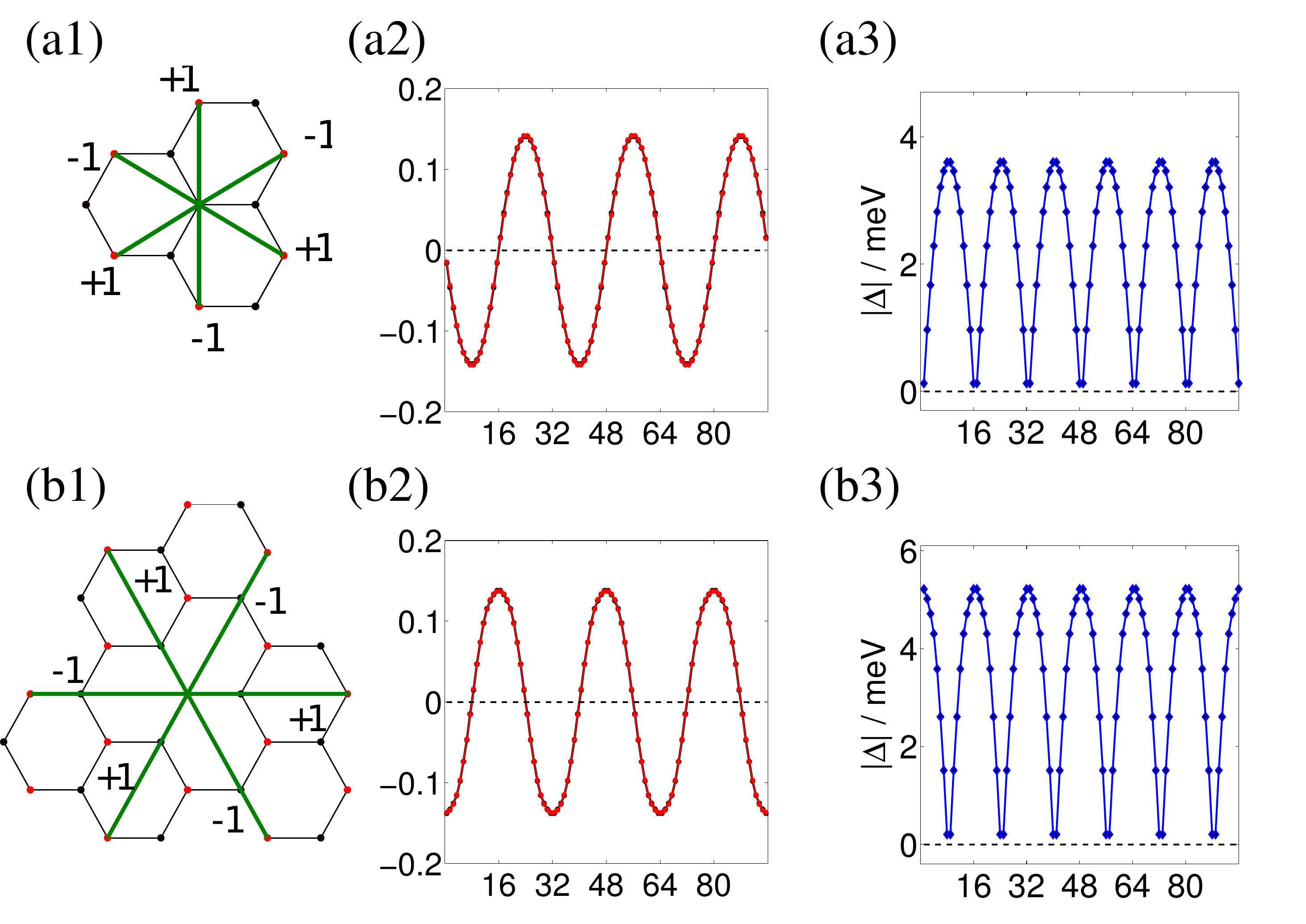}
  \end{minipage}
  \caption{(Color online). Pairing amplitudes, form factors, and gap profiles for the $f$-wave phase as defined in Fig.~\ref{pic3}, representative for larger fillings than the VHS for (a) $U_1/U_0=0.5$ to (b) $U_1/U_0=0.5, U_2/U_==0.3$. The gap profile is nodal, the nodal points shift from (a) to (b).}
\label{pic4}
\vspace{-0pt}
\end{figure}

{\it $f$-wave phase.} It is likewise interesting to investigate the triplet $f$-wave instability~\cite{sri} which dominates for longer ranged Coulomb interaction [Fig.~\ref{pic4}]. It obeys the one dimensional $B_1$ or $B_2$ representation, depending on the range of the Coulomb interaction. 
For $U_1/U_0=0.5$, the form factor and pairing amplitudes are plotted in Fig.~\ref{pic4}a as well as for $U_1/U_0=0.5$, $U_2/U_0=0.3$ in Fig.~\ref{pic4}b. We again find that the Cooper pair distance increases, which manifests itself as a change of the form factor $f[f_{B_1}]=\sin(\sqrt{3}k_y) - 2\sin(\sqrt{3}k_y/2)\cos(3k_x/2)$ in (a) to $f[f_{B_2}]=\sin(3k_x) - 2\sin(3k_x/2)\cos(3\sqrt{3}k_y/2)$ in (b). The gap function follows the absolute value of the form factor, showing a nodal gap, where the points of the nodes change with increasing Coulomb range. In the case of $f$-wave, the position of the nodes would hence indicate the Cooper pairing distance associated with the long range properties of the Coulomb interaction, and suggest experimental evidence of a nodal gap from transport and an invariant Knight shift due to triplet pairing. For filling smaller than the VHS, the Fermi surface is disconnected and it can happen that the nodes do not coincide with the Fermi surfaces. While probably very low in $T_c$, depending on whether $B_1$ or $B_2$ is preferred, this $f$-wave regime could be nodeless.

{\it Summary and outlook.} In summary, we have provided a detailed analysis of the competing many-body phases of graphene at and around van Hove filling. We find that for realistic band structure parameters and interactions, the exotic nodeless singlet $d+id$-wave SC phase appears to be preferred over an extended phase space regime around the VHS. Variations of the kinetic parameters and effective interaction scales can drive a transition to a spin density wave phase at the van Hove point. In a broader vicinity of the VHS, reduced Coulomb screening changes the form of the $d+id$ Cooper pair wave function, and in certain limits might favor a nodal triplet $f$-wave SC phase.

The possibility of the time-reversal symmetry breaking $d+id$ phase in graphene is very intriguing: it has been noted early on in the context of the cuprates that such a phase would have various interesting properties such as quantized edge currents~\cite{laughlin-dpid,sigrist}. Furthermore, provided Rashba spin-orbit interaction is present, $d+id$ phase supports Majorana modes in the vortex cores obeying non-Abelian statistics~\cite{sato10prb82}.
The tunability of the Rashba interaction in graphene~\cite{min06prb74} may enable realization of the Majorana modes; owing to the two-dimensional nature of graphene and its remarkable tunability, their observation and manipulation should be easier than in other materials.

\begin{acknowledgments}
  RT thanks A. Chubukov, C.~Honerkamp, G. Baskaran, S.~Raghu, and L.~Boeri for discussions.
  CP, WH, and RT are supported by
  DFG-SPP 1458/1, CP by DFG-FOR 538. RT is supported by an SITP fellowship of Stanford University.
\end{acknowledgments}

\bibliographystyle{prsty}
\bibliography{bibliography_new}

\end{document}